\documentclass[twocolumn,floatfix,superscriptaddress]{revtex4-2}
\usepackage{dsfont} 
\usepackage{bm} 

\usepackage{bbold} 
\usepackage{xcolor}
\usepackage{times}
\usepackage{float}
\usepackage{physics}
\usepackage{graphicx,braket}
\usepackage{hyperref}
\hypersetup{colorlinks=true,linkcolor=blue,citecolor=blue}
\usepackage{amsmath,amssymb,amsfonts}
\usepackage{wrapfig}

\usepackage{cleveref} 

\usepackage{siunitx}

\def\ket#1{\mathinner{|{#1}\rangle}} 

\usepackage{cleveref}
\crefname{equation}{Eqs.}{Eqs.}
\Crefname{equation}{Equation}{Equations}
\crefrangelabelformat{equation}{(#3#1#4--#5#2#6)}
\crefmultiformat{equation}{Eqs. (#2#1#3}{, #2#1#3)}{#2#1#3}{#2#1#3}
\Crefmultiformat{equation}{Equations (#2#1#3}{, #2#1#3)}{#2#1#3}{#2#1#3}
\usepackage{ amssymb }
\usepackage{lipsum}
\usepackage{kantlipsum}

\begin{document}
	
	\allowdisplaybreaks 
	
	\flushbottom
	\title{Unconventional mechanism of virtual-state population through dissipation}
	\author{Alejandro Vivas-Via{\~n}a}
	\affiliation{Departamento de Física Teórica de la Materia Condensada and Condensed
		Matter Physics Center (IFIMAC), Universidad Autónoma de Madrid, 28049 Madrid,
		Spain}
	\author{Alejandro González-Tudela}	
	\affiliation{Institute of Fundamental Physics IFF-CSIC, Calle Serrano 113b, 28006 Madrid, Spain}	
	\author{Carlos S\'anchez Mu\~noz}
	\email[]{carlos.sanchezmunnoz@uam.es}
	\affiliation{Departamento de Física Teórica de la Materia Condensada and Condensed
		Matter Physics Center (IFIMAC), Universidad Autónoma de Madrid, 28049 Madrid,
		Spain}
	
	\newcommand{\down}{\op{g}{e}}
	\newcommand{\up}{\op{e}{g}}
	\newcommand{\downd}{\op{+}{-}} 
	\newcommand{\upd}{\op{+}{-}}
	\newcommand{\app}{a^\dagger}
	\newcommand{\ssp}{\sigma^\dagger}
	\newcommand*{\Resize}[2]{\resizebox{#1}{!}{$#2$}}%

	\begin{abstract}
	Virtual states are a central concept in quantum mechanics. By definition, the probability of finding a quantum system in a virtual state should be vanishingly small at all times. In contrast to this notion, we report a phenomenon occurring in open quantum systems by which virtual states can acquire a sizable population in the long time limit, even if they are not directly coupled to any dissipative channel. This means that the situation where the virtual state remains unpopulated can be metastable.
We describe this effect by introducing a two-step adiabiatic elimination method, that we termed hierarchical adiabatic elimination, which allows one to obtain analytical expressions of the timescale of metastability in general open quantum systems.
We show how these results can be relevant for practical questions such as the generation of stable and metastable entangled states in dissipative systems of interacting qubits.

	\end{abstract}
	\date{\today} \maketitle
	
	\textit{Introduction.}--- The concept of virtual in quantum mechanics is of paramount importance, e.g. in the context of virtual transitions between coherently unconnected states~\cite{CohenTannoudji1997,CohenTannoudji1998,Sakurai2017} or in the description of scattering processes in QFT where interactions are mediated by virtual particles~\cite{Ryder1996}. In situations where strongly off-resonant ``virtual'' states mediate interactions between quasi-resonant ``real'' states, an adiabatic elimination over the fast degrees of freedom---the virtual ones---allows one to reduce the dimensionality of the problem and obtain an effective description of the slow degrees of freedom, i.e. the real states. This technique of adiabatic elimination, which can be formulated in several alternatives ways---e.g., the Schrieffer-Wolff transformation~\cite{CohenTannoudji1998}---is ubiquitous in the description and design of quantum phenomena, e.g.  quantum optical applications in atomic physics~\cite{Gaubatz1990,Bergmann1998,Lutkenhaus1998,
Warszawski2000,Brion2007,Dimer2007,Burgarth2019,Gamel2010,Damanet2019,Kaufman2020,Burgarth2021} or exotic dynamics in the ultrastrong coupling regime of cavity QED~\cite{Garziano2015,Garziano2016,Stassi2017}. A significant effort has been made to establish the mathematical foundations of this technique~\cite{Comparat2009,Mirrahimi2009,Paulisch2014} and its extension to dissipative contexts for its application in open quantum systems~\cite{Santos2021,Reiter2012,Azouit2016,Finkelstein-Shapiro2020}. 
	
The fundamental underlying assumption for the adiabatic elimination of a virtual state is that the coupling between the real subspace $\mathcal H_R$ and the virtual subspace $\mathcal H_V$ is perturbative;  i.e., the coherent coupling rate is much smaller than the energy difference between subspaces; as a result, one can obtain an effective Hamiltonian acting only in $\mathcal H_R$. Consequently, when this approximation applies, any initial state in  $\mathcal H_R$ will remain within that subspace, and $\mathcal H_V$ will not be populated. In this work, we show that the situation can be radically different when there is also dissipative dynamics, even if dissipative process only take place within $\mathcal H_R$. We unveil an unconventional mechanism by which, in the long time limit,  virtual states acquire a sizable occupation probability, comparable to that of the real states. These findings can have great importance in the understanding and engineering of interactions between quantum systems in driven-dissipative contexts~\cite{Plenio1999,Diehl2008,Verstraete2009,Chang2018}.

In order to study this phenomenon, we start discussing what is arguably the simplest scenario that can be described in terms of virtual states [see Fig.{\ref{fig:fig1_setup}(a)]: two quasi-resonant “real” states, effectively interacting through the mediation of a third, strongly off-resonant “virtual” state. Crucially, we enable a spontaneous decay between the real states, which can be provided, for instance, by the coupling to a surrounding environment in a Markovian regime. Contrary to the familiar intuition, the situation in which the virtual state remains ``virtual'' is, in this case, only metastable~\cite{Macieszczak2016,Macieszczak2021}, and, in the long time limit, the system eventually relaxes to a stationary state where the virtual state has a sizable population.

This process of \emph{de-virtualization} occurs through an unconventional mechanism of population enabled by dissipation. In this work, we introduce a technique of \emph{hierarchical adiabatic elimination} to obtain analytical approximations of the time-dependent elements of the system density matrix and expressions for the characteristic metastability timescales.
We show how our novel technique can be used to described metastable dynamics in different systems involving two interacting qubits, where the phenomenon reported has strong implications for the generation of stable and metastable entanglement via dissipation.

	\textit{Model.}--- The first model we study consists of a three-level system configuration, sketched in Fig.~\ref{fig:fig1_setup}(a). The Hilbert space spans a basis $\left\{|1\rangle , |2\rangle, |V\rangle \right\}$, where the states $|1\rangle$ and $|2\rangle$ represent two real states, and $|V\rangle$ will play the role of a virtual state, being strongly detuned from $|1\rangle$ and $|2\rangle$. 	We define lowering operators as $\hat \sigma_{i,j}\equiv |i\rangle \langle j|$  $(i,j ={1,2,V})$. The real states are coupled to $|V\rangle$ with a coupling rate $\Omega$, and there is an irreversible decay process within the real subspace, with state $|2\rangle$ decaying towards $|1\rangle$ with a decay rate $\Gamma$. This specific $\Lambda$ model could be motivated, for instance, by the description of a quantum-optical system consisting of two interacting qubits coherently excited at the two-photon resonance with a Rabi frequency $\Omega$~\cite{Varada1992,Hettich2002,Haakh2015,Vivas-Viana2021}, where the ground state $|gg\rangle$ corresponds to the real state $|1\rangle$, the doubly-excited state $|ee\rangle$ is the excited real state $|2\rangle$, and the symmetric single-excitation state $|S\rangle = \frac{1}{\sqrt{2}}(|eg\rangle + |ge\rangle)$---detuned from the two-photon transition energy due to the interaction between qubits---corresponds to the virtual state $|V\rangle$. The two-photon decay channel can be enabled, for instance, by a cavity in resonance with the two-photon transition~\cite{DelValle2010a,Ota2011}. A change to the rotating frame of the drive would directly yield the configuration shown in Fig.~\ref{fig:fig1_setup}(a).	
 	\begin{figure}[t!]
		\includegraphics[width=0.45\textwidth]{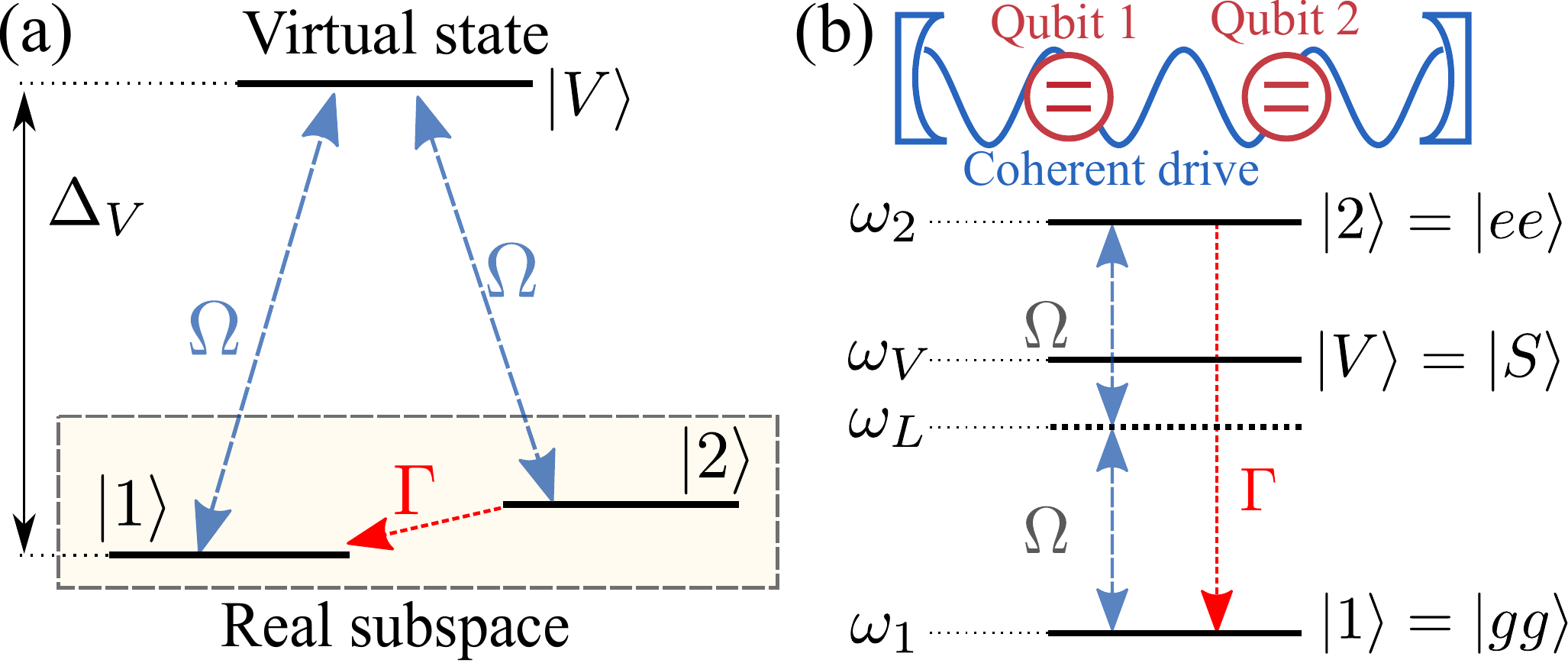}
		\caption{a) Scheme of the system: two quasi-resonant ``real'' quantum states, interacting via a third, strongly off-resonant “virtual” state. There is spontaneous decay between the real states. (b) The system in (a) can describe two interacting two-level systems under coherent driving at the two-photon resonance, in the rotating frame of the drive. Decay between real states can be engineered with a cavity.}
		\label{fig:fig1_setup}
	\end{figure}
The resulting time-independent Hamiltonian is $\hat H=\hat H_0+\hat H_d$, where $\hat H_0$ is the bare Hamiltonian ($\hbar=1$),
		\begin{equation}
		\hat H_0=\Delta_1 |1\rangle \langle 1  |+ \Delta_2 |2\rangle \langle 2 | + \Delta_V |V\rangle \langle V|, 
	\end{equation}
	and $\hat H_d$ is the Hamiltonian of the driving/coupling term
	\begin{equation}
		\hat H_d=\Omega \left(\hat \sigma_{1,V}+\hat \sigma_{2,V}+ \text{H.c}  \right), 
	\end{equation}
	where $\Delta_i$ ($i=1,2,V$) stand for the free energy parameters, where we will assume that $\Delta_V\gg \Delta_2,\Delta_1,\Omega$ and $\Delta_2\approx \Delta_1 \approx 0$. We assume that the evolution of the system is governed by a quantum master equation~\cite{Breuer2007},
	\begin{equation}
		\frac{d \hat\rho}{dt}= -i[\hat H, \hat \rho]+\frac{\Gamma}{2} \mathcal{L}_{\hat \sigma_{21}}[\hat \rho ] + \frac{\Gamma_V}{2} \mathcal{L}_{\hat \sigma_{1V}}[\hat \rho ],
		\label{eq_Master}
	\end{equation}
	where the Lindblad term $ \mathcal{L}_{\hat O} \equiv 2 {\hat O} \hat \rho {\hat O}^\dagger-\left\{{\hat O}^\dagger {\hat O},\hat \rho\right\}$ describes processes of spontaneous decay. Unless stated otherwise, we will consider $\Gamma_V=0$, i.e.,  we assume there is only one process of spontaneous decay, from $|2\rangle$ to $|1\rangle$ (the case $\Gamma_V\neq 0$ will be considered later only for comparison). 
The dynamics of the system can be studied straightforwardly by numerically solving Eq.~\eqref{eq_Master}. Figure~\ref{fig:fig2_dynamics}(a) shows the occupation probability of the excited state, $\rho_{2,2}\equiv \langle 2|\hat\rho|2\rangle$ and the virtual state $\rho_{V,V} \equiv \langle V|\hat\rho| V\rangle$ versus time. One can clearly appreciate the existence of two distinct relaxation timescales.  Within the first relaxation timescale ($t \sim 1/\Gamma$),   the system behaves according to the standard intuition regarding virtual states: $|V\rangle$ remains unpopulated, mediating the interaction between $|1\rangle$ and $|2\rangle$, which gives rise to coherent Rabi oscillations between these two states with a two-photon Rabi frequency $\Omega_{2\mathrm p} =  \Omega^2/\Delta_V$, damped by spontaneous emission of rate $\Gamma$ into a stationary state. This situation can be described simply in terms of a coherently driven two-level system spanned by $|1\rangle$ and $|2\rangle$. This stationary regime is, however, metastable, and in a much longer timescale, which for this particular choice or parameters is $t\sim 10^4/\Gamma$, $\rho_{V,V}$ develops a population comparable to $\rho_{2,2}$. Clear evidences of this metastable behaviour in open quantum systems can be found in the spectrum of eigenvalues  of the Liouvillian superoperator $\mathcal L$~\cite{Macieszczak2016,Macieszczak2021}. All these eigenvalues  $\{\lambda_k, k=1,2,\ldots \}$---ordered here by its real values, so that $\Re(\lambda_k)\geq \Re(\lambda_{k+1})$---have a negative real part, and the eigenvalue with the largest real part is necessarily equal to zero, $\lambda_1 =0$, its corresponding eigenstate being the steady state of the system.  The second largest real value of the Liouvillian spectrum, $\Re(\lambda_2)$, is the Liouvillian gap~\cite{Kessler2012}, and it gives the relaxation time necessary to reach the steady state, $\tau_2 = 1/|\Re(\lambda_2)|$. Metastability results when $\lambda_2$ is well separated from the rest of eigenvalues by a second gap, so that $\Re(\lambda_3)\ll \Re(\lambda_2)$~\cite{Macieszczak2016} (here, we assume for simplicity that, as in the case of our model, there is only one metastable state, rather than a manifold). Then, the system relaxes to a metastable state in a timescale $\tau_3 = 1/|\Re(\lambda_3)|$, which will eventually evolve into the actual steady state in a time $\tau_2 \gg \tau_3$.  The system we consider here exhibits precisely this clustering of eigenvalues characteristic of metastability, as can be seen in Fig.~\ref{fig:fig2_dynamics}(b) where we confirm that $\tau_3 \sim 1/\Gamma$, and $\tau_2 \sim 10^4/\Gamma$. 
The steady state value of the virtual state occupation probability can be computed analytically, yielding
	\begin{equation}
		\rho_{V,V}^{\mathrm{ss}}=\frac{\Omega ^2 \left(\Gamma ^2+4 \Omega ^2\right)}{2 \Omega ^2 \left(\Gamma ^2+6 \Omega ^2\right)+\Gamma ^2 \Delta _V^2}.
	\end{equation}
In the limit $\Omega^2 \gg \Gamma \Delta_V$, this expression indeed yields a sizable population $\rho_{V,V}^{\mathrm{ss}}\approx 1/3$, clearly establishing that the virtual state will get populated in the long time limit. 
	
	\begin{figure*}[t!]
		\begin{center}
			\includegraphics[width=0.98\textwidth]{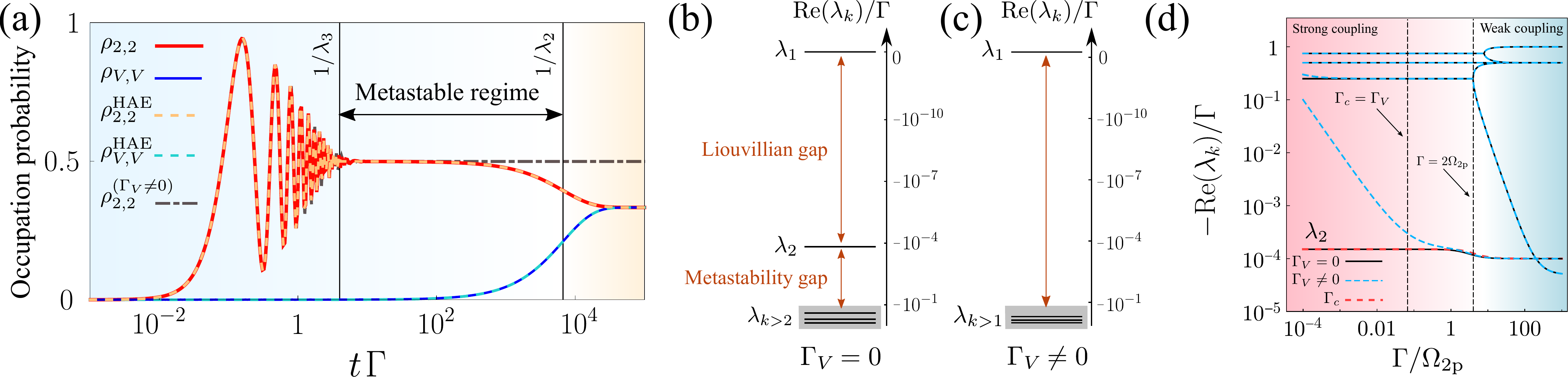}
		\end{center}
		\caption{(a) Dynamics of the excited real state and the virtual state. Solid lines are numerical calculations, dashed lines are analytical predictions from the hierarchical adiabatic elimination. Dot-dashed represents a case where $\Gamma_V\neq 0$, showing the stabilization of the metastable state. (b) Liouvillian spectrum, featuring the defining characteristic of metastability in open quantum systems: a metastability gap between $\lambda_2$ and  $\lambda_{k>2}$.  (c) Same as in (b), with $\Gamma_V= \Gamma$. In this system, metastability is no longer present. (d) Liouvillian eigenvalues versus $\Gamma$, for $\Gamma_V=0$ (solid, black) and $\Gamma_V = 10^{-5}\Omega_{2\mathrm p}$. Our analytical prediction for the value of the Liouvillian gap $\lambda_2$ for $\Gamma_V=$ is shown in dashed-red. If $\Gamma_V \neq 0$, it is seen that metastability disappears for when $\Gamma_c < \Gamma_V$. Parameters: $\Omega/\Delta_V=0.01$; in (a---c), $ \Gamma/\Delta_V=10^{-5}$. $\Gamma_V$ is zero unless indicated otherwise, in which case $\Gamma_V = \Gamma$. }
		\label{fig:fig2_dynamics}
	\end{figure*}
	
		\textit{Hierarchical Adiabatic elimination (HAE)}.---In order to have an estimate of the survival time of the metastable state, it would be desirable to obtain an analytical expression of $\lambda_2$. A direct analytical solution for the time evolution of the density matrix through the diagonalization of $\mathcal L$ is not readily available. Nevertheless, it is clear from our previous discussion the existence of a hierarchy of timescales, which suggests that a series of adiabatic elimination techniques could be applied. \emph{(i)} The shortest timescale is clearly governed by Hamiltonian dynamics, evidenced by a fast oscillatory evolution of the density matrix elements. This oscillatory dynamics stabilizes into a steady state in a timescale of the order $\tau_3 \sim 1/\Gamma$. In these timescales, the population of $|V\rangle$ plays the role of a fast variable: $|V\rangle$ mediates effective interactions within the real subspace, i.e., it plays the role of a virtual state that can be eliminated within a purely Hamiltonian evolution. This is the first adiabatic elimination that we will perform. \emph{(ii)} The longest timescale is characterized by a very slow evolution of $|V\rangle$. In this long timescale, the relaxation of the real variables in a time $1/\Gamma$ occurs almost instantaneously, meaning that one can treat the real variables as the \emph{fast} variables in a dissipative sense, i.e., they relax quickly into a time-dependent quasi-steady state that follows the slow evolution of $\rho_{V,V}$. This can be described in terms of a second adiabatic elimination. Note that a direct application of standard adiabatic elimination techniques in dissipative context, e.g. the projection-operator method~\cite{Haken2004,Finkelstein-Shapiro2020}, would directly eliminate the virtual subspace and, therefore, it would fail to capture the mechanism that populates this state in the long time limit. We present instead a two-step method, that we label hierarchical adiabatic elimination.
		
\textit{First adiabatic elimination}.---Our starting point is the set of differential equations describing the evolution of the elements of the total density matrix, obtained from Eq.~\eqref{eq_Master} as
\begin{subequations}
		\begin{align}
\label{eq:drho_VV}
			\dot \rho_{V,V}&=
			2 \Omega \Im\left[ \rho_{1,V}-\rho_{V,2}\right], \\
			\dot \rho_{2,2}&=-\Gamma \rho_{2,2}+2\Omega \Im \left[\rho_{V,2}\right], \\
			\dot \rho_{1,V}&=i \Delta_V \rho_{1,V}
			 +i\Omega \left[1+\rho_{1,2}-2\rho_{V,V}-\rho_{2,2}\right], \\
			\dot \rho_{1,2}&=-(\Gamma/2)\rho_{1,2}+i\Omega \left[\rho_{1,V}-\rho_{V,2}\right], \\
			\dot \rho_{V,2}&=-\left(i \Delta_V+\Gamma/2\right)\rho_{V,2}			-i\Omega \left[\rho_{1,2}+\rho_{2,2}-\rho_{V,V}\right].
		\end{align}
\end{subequations}
Based on the assumption that $|\Delta_V - \Delta_i |\gg |\Omega|$ ($i\in \{1,2\}$), we perform an adiabatic elimination consisting in setting $\dot\rho_{1,V}=\dot\rho_{V,2} =0$. In the limit $\Delta_V\gg \Gamma$, the resulting effective equations that govern the dynamics of the real subspace become
\begin{subequations}
\begin{align}
\label{eq:adiabatic-1st-a}
\dot\rho_{2,2} &\approx -\Gamma \rho_{2,2} -2 \Omega_{2\mathrm p}\, \text{Im}[\rho_{1,2}] + \frac{\Gamma \,\Omega_{2\mathrm p}}{\Delta_V}\, \rho_{V,V},\\
\dot\rho_{1,2} &\approx-\Gamma/2\, \rho_{1,2}+i\Omega_{2\mathrm p}( 2\rho_{2,2}+\rho_{V,V} -1),
\label{eq:adiabatic-1st-b}
\end{align}
\end{subequations}
where we defined a two-photon Rabi frequency, $\Omega_{2\mathrm p }\equiv \Omega^2/\Delta_V$. 
These equations can be solved  considering $\rho_{V,V}$ as a time-independent parameter with a fixed value (i.e. $\dot\rho_{V,V}=0$). A natural choice would be to set $\rho_{V,V}=0$. In that case, \cref{eq:adiabatic-1st-a,eq:adiabatic-1st-b} simply describe the dynamics of two resonant levels coupled  via a second-order process with a Rabi frequency of $\Omega_{2\mathrm p}$ with a standard decay; e.g., the  regime of  coherent two-photon driving of the transition $|gg\rangle \leftrightarrow |ee\rangle$ in the case of a two-atom system depicted in Fig.\ref{fig:fig1_setup}(b). Such a two-level system dynamics describes accurately the short-timescale oscillatory dynamics of Fig.~\ref{fig:fig2_dynamics}(a), where the initial state was set to be $|1\rangle$. From now on, we focus on the strong coupling limit $\Gamma \lesssim\Omega_{2\mathrm p}$; otherwise, the system is  overdamped and will basically remain in the ground state $|1\rangle$. 

\emph{Second adiabatic elimination---} While the usual approach when eliminating a virtual state is to indeed  assume $\rho_{V,V}=0$ for all times, we have already seen that this approach is eventually bound to fail, since $\rho_{V,V}$ develops a sizable population within a characteristic timescale $\tau_2\gg 1/\Gamma$ which, crucially, is orders of magnitude longer than the relaxation time of \cref{eq:adiabatic-1st-a,eq:adiabatic-1st-b}. This suggest we can make a second adiabatic elimination based on this separation of timescales. From the first adiabatic elimination conditions ($\dot\rho_{V,i}=0$) and Eq.~\eqref{eq:drho_VV}, we can obtain a differential equation for $\rho_{V,V}$ which is a function of itself and the real-subspace elements, i.e.  $\dot\rho_{V,V}(t) = f[\rho_{V,V}(t);\rho_{1,2}(t);\rho_{2,2}(t)]$ (see Supplemental Material for a full expression). The second adiabatic elimination consist then in substituting $\rho_{1,2}(t)$ and  $\rho_{2,2}(t)$ in that equation by their steady state solutions of \cref{eq:adiabatic-1st-a,eq:adiabatic-1st-b} obtained for a given $\rho_{V,V}=\rho_{V,V}(t)$, yielding a dynamical equation that only depends on $\rho_{V,V}(t)$, i.e. $\dot\rho_{V,V}(t) = f[\rho_{V,V}(t);\rho_{1,2}^{\mathrm{ss}}(\rho_{V,V}(t));\rho_{2,2}^{\mathrm{ss}}(\rho_{V,V}(t))]$. 
Here, $\rho_{2,2}$ and $\rho_{1,2}$ act as fast variables that relax into a time-dependent stationary state that follows the slow evolution of $\rho_{V,V}$. Solving this differential equation, one obtains ${\rho_{V,V}(t)\approx \rho_{V,V}^{\mathrm{ss}}\left(1-e^{-\Gamma_c t}\right)}$, where we have defined the relaxation rate
	\begin{equation}
\Gamma_c \approx \frac{3\Gamma \Omega^2}{2\Delta_V^2},
\label{eq:Gamma_c}
	\end{equation}
obtained under the assumption  $\Omega_{2\mathrm p} \gg \Gamma $  (a full, more cumbersome expression that does not require that assumption is provided in the Supplementary Material). 	Equation~\eqref{eq:Gamma_c} is the desired expression that gives us the survival time of the metastable regime, and thus, it must correspond to the Liouvillian gap, $\Gamma_c = |\Re(\lambda_2)|$. We have checked that this is indeed the case in Fig.~\ref{fig:fig2_dynamics}(d), which depicts the spectrum of eigenvalues of $\mathcal L$ as a function of $\Gamma$, showing a perfect match between our analytical expression of $\Gamma_c$ and $\lambda_2$.  We also note the perfect matching of the analytical solutions of the dynamics in Fig.~\ref{fig:fig2_dynamics}(a) (full expressions in Supplemental Material). 
	
	\textit{Mechanism of population}.---We will now provide some insights into the dissipative mechanism that results in the population of the virtual state.  	 In order to do so, we perform an analysis from the perspective of quantum trajectories using the method of quantum jumps~\cite{Plenio1998,Brun2002,Gerry2004}. 	Inspection of individual trajectories---see e.g. the example of Fig.~\ref{fig:fig3_nojump}(a)--- reveals that the virtual state gets populated through non-Hermitian evolution between quantum jumps---the effect of a jump is, in fact, to strongly decrease the population of the virtual state---. This can be understood if one considers the information about the system leaked to the environment during a time interval with no jumps~\cite{haroche_book06a}. When no jump occurs, the system is logically more likely to be in a state that cannot emit, i.e., either $|1\rangle$ or $|V\rangle$. However, since $|1\rangle$ is resonantly coupled to $|2\rangle$ to second-order in perturbation theory, a system in $|1\rangle$ will eventually evolve into $|2\rangle$ and lead to a jump. In other words, as a period without a jump becomes longer, $|V\rangle$ becomes the most likely state, and the system is updated accordingly, increasing its population.  	This purely-dissipative mechanism will slowly accumulate over time, explaining the population buildup of the virtual state in our system. This intuition is further confirmed by computing a conditional density matrix for the particular trajectory in which no jumps occur at all, see Fig.~\ref{fig:fig3_nojump}(b). In this particular scenario (whose probability naturally decreases over time), the population of the virtual state saturates to its maximum possible value.
		\begin{figure}[t!]
		\centering
		\includegraphics[width=0.45\textwidth]{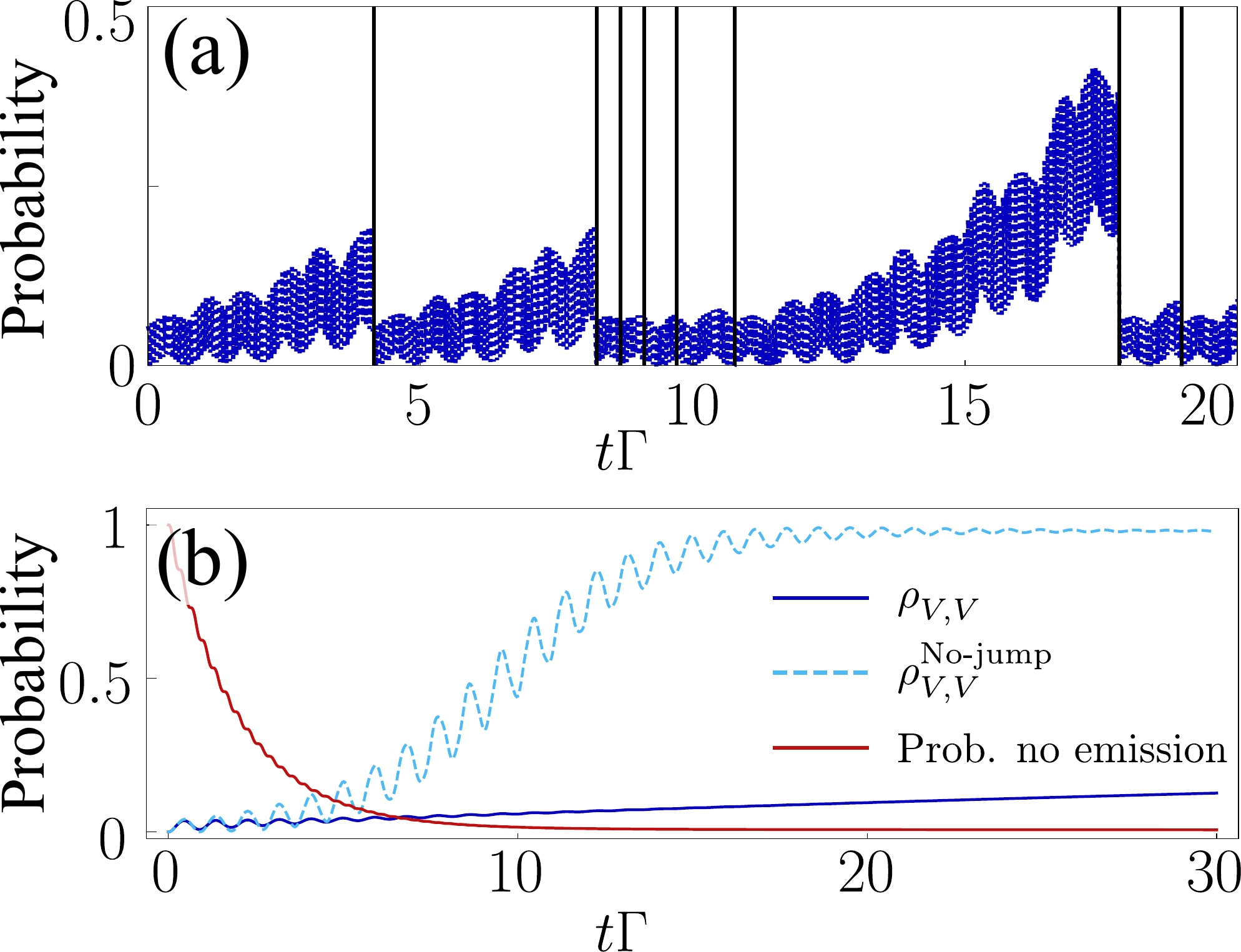}
		\caption{(a) Population of the virtual state through non-Hermitian evolution between quantum jumps. (b)  Conditional evolution when no jumps take place. Red line: probability of no-jump. Blue dashed line: population of the virtual state conditioned to no jumps, showing that, in this case, it reaches its maximum possible value. Blue straight line: population of the virtual state for the general evolution. 
			Parameters: $\Omega/\Delta_V=0.1,\ \Gamma/\Delta_V=10^{-3},\  \Delta_2=\Delta_1=0.$}
		\label{fig:fig3_nojump}
	\end{figure}

This mechanism would be completely disrupted if there were additional dissipative channels involving the virtual state. We can consider this situation by setting $\Gamma_V\neq 0$ in Eq.~\eqref{eq_Master}, thus including a channel of spontaneous emission from the virtual state to the ground state. The ratio between decay rates, $\Gamma_V/\Gamma_c$, will determine whether virtual state population occurs or not. When $\Gamma_V \gtrsim \Gamma_c$, dissipation from $|V\rangle$ outcompetes the mechanism of population, and one recover the simple dynamics in terms of a driven two-level system, as can be seen in Fig.~\ref{fig:fig2_dynamics}(a).  Consistently with this, there is no longer a Liouvillian eigenvalue corresponding to a metastable state---as shown in  Fig.~\ref{fig:fig2_dynamics}(c-d)---, as $\Gamma_V$ becomes comparable to $\Gamma_c$, $\lambda_2$ is pulled towards values $\sim \Gamma$ and the metastability gap disappears. 

\textit{Generality of the HAE and its implications for entanglement generation.}---The HAE method introduced in this work can be a valuable tool to obtain analytical insights about metastable dynamics in open quantum sytems. This can have strong implications for quantum technological applications such as the generation of entanglement in dissipative quantum systems. In order to illustrate this, we now apply the HAE to describe entanglement generation in  two different systems displaying metastability. 

First, we consider entanglement generation in the system of two interacting qubits already described in Fig.~\ref{fig:fig1_setup}(b), which, as noted before, maps into the the three level model discussed so far in this text, assuming that the occupation of the antisymmetric state $|A\rangle \propto (|eg\rangle - |eg\rangle)$ is  completely decoupled from the dynamics and remains equal to zero. This map between models allows us to use the density matrix elements estimated with the HAE---c.f. Fig.~\ref{fig:fig2_dynamics}(a)---to compute the concurrence and  quantify the degreee of entanglement between the two qubits~\cite{Wootters1998,Wootters2001,Plenio2007,Horodecki2009}.
 The results are shown in Fig.~\ref{fig:fig4_ConcurrencePlot}(a), evidencing the formation and stabilization of entanglement at short timescales $t\sim 1/\Gamma$, due to the coherence built between the states $|ee\rangle$ and $|gg\rangle$ via the two-photon drive. Notably, this entanglement is long-lived, but metastable, and its survival time is given by $1/\Gamma_c$, i.e., the relaxation rate obtained in Eq.~\eqref{eq:Gamma_c} via the HAE. The unconventional population of the virtual state $|S\rangle$ thus destroys entanglement in the long time limit. 
 
Next, we consider another two-qubit system, sketched in the inset of Fig.~\ref{fig:fig4_ConcurrencePlot}(b). In this case, qubits experience collective decay with rate $\Gamma$, inducing transitions $|ee\rangle \rightarrow |S\rangle$ and $|S\rangle \rightarrow |gg\rangle$. Furthermore, each qubit is driven with the same Rabi frequency $\Omega$, but each of them is detuned from the drive frequency by an absolute value $\delta$ with opposite sign. This system was introduced in Refs.~\cite{Ramos2014,Pichler2015} in the context of chiral waveguides; the Hamiltonian is $\hat H = \sqrt{2}\Omega (|S\rangle\langle gg| + |ee\rangle \langle S| ) + (\delta-i\Delta\gamma/2)|A\rangle\langle S| + \mathrm{H.c.}$, with $\Gamma = 2(\gamma_R + \gamma_L)$ and $\Delta\gamma \equiv \gamma_R-\gamma_L$, where $\gamma_R$ and $\gamma_L$ describe decay into right and left propagating modes respectively. This configuration was shown to stabilize in the long time limit to the fully-entangled dark state $|A\rangle$ provided $\Omega \gg \Delta\gamma, \delta$, as we show explicitly in Fig.~\ref{fig:fig4_ConcurrencePlot}(b). The application of HAE to describe this system follows exactly the same reasoning detailed above, with $|A\rangle$ playing the role of the ``virtual'' state that gets populated over time. Using this technique, we are able to establish the timescale of formation of the entangled state, which is given by $\tau \approx 24\Omega^2/[\Gamma(4\delta^2+\Delta\gamma^2)]$. The analytical results obtained from the HAE method match perfectly the exact calculations, as shown in Fig.~\ref{fig:fig4_ConcurrencePlot}(b). In the limit $(\delta,\Delta\gamma)\rightarrow 0$ we find $\tau\rightarrow \infty$, meaning that the metastable state becomes the steady state, as reported, e.g., in Ref~\cite{Gonzalez-Tudela2011}.

	\begin{figure}[t]
		\centering
		\includegraphics[width=0.45\textwidth]{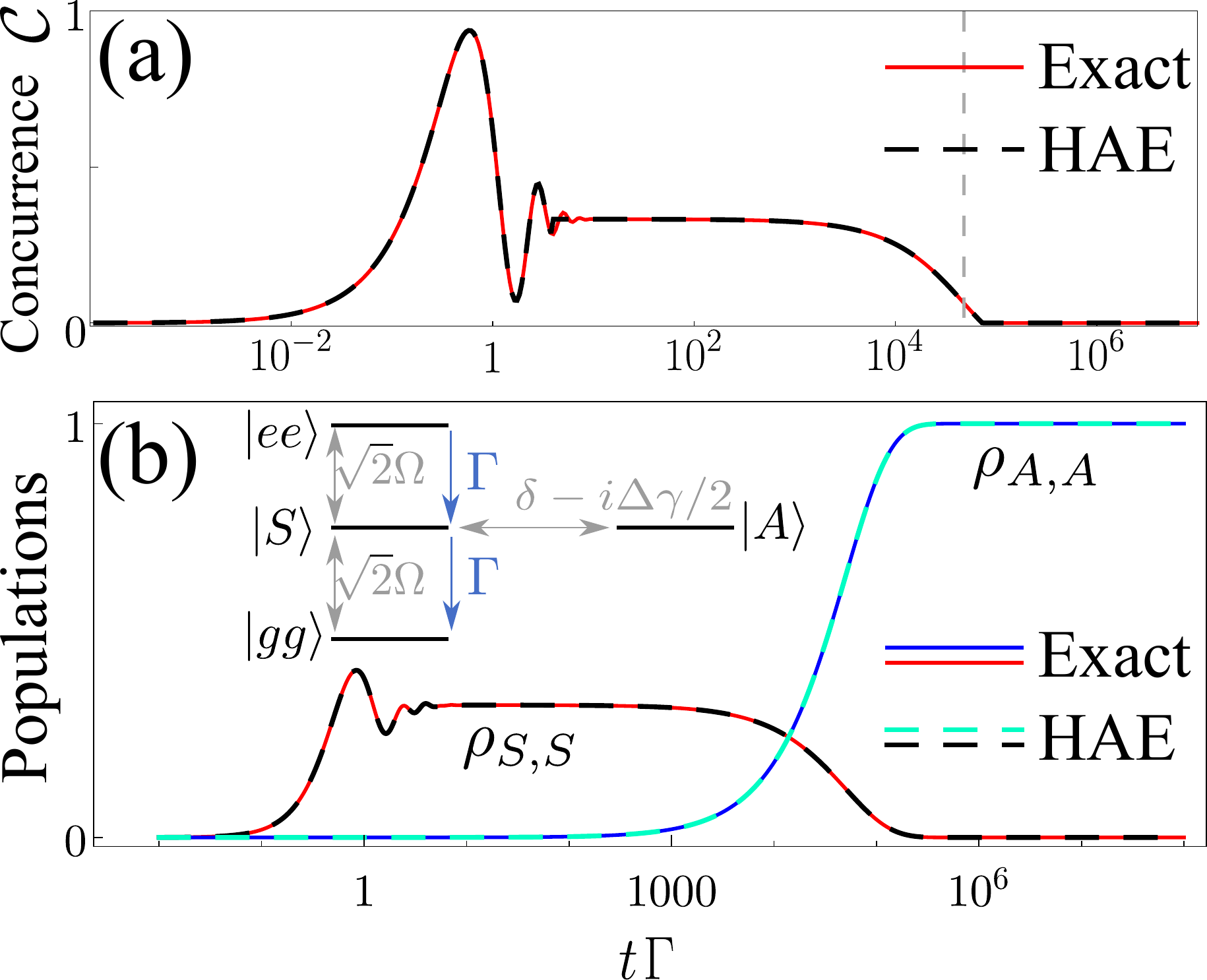}
		\caption{
			 Entanglement formation in systems of two coupled qubits. (a) Formation of metastable entangled state in the system sketched in Fig.~\ref{fig:fig1_setup}(b), surviving for a time $\tau_2 \sim 1/\Gamma_c$. 			 
			Parameters: $\Gamma/\Delta_V=10^{-5},\  \Delta_2/\Delta_V=0.$ (b) Stabilization of the entangled antisymmetric state in the two-qubit system sketched in the inset. The evolution is perfectly described by the HAE. Parameters: $\Omega/\Gamma=1$, $\delta/\Gamma=0.01$, $\Delta\gamma/\Gamma=0.01$.  }
		\label{fig:fig4_ConcurrencePlot}
	\end{figure}

\emph{Conclusion---.} We have shown that, in open quantum systems, off-resonant virtual states can get populated in the long time limit even if they are not connected to any dissipative channel, meaning that the regime where the virtual state is not populated is metastable. We introduce a method of hierarchical adiabatic elimination that approximates the dynamics and provides analytical expressions of the lifetime of the metastable state. Our method can be applied in a variety of metastable open quantum systems to obtain valuable insights in questions such as the dissipative stabilization of entangled states.

\acknowledgements
	The authors are thankful to C. Navarrete-Benlloch for insightful discussions. C.S.M. acknowledges that the project that gave rise to these results received the support of a fellowship from la Caixa Foundation (ID 100010434), from the European Union's Horizon 2020 Research and Innovation Programme under the Marie Sklodowska-Curie Grant Agreement No. 847648, with fellowship code  LCF/BQ/PI20/11760026. A.G.T. acknowledges support from  CSIC Research   Platform   on   Quantum   Technologies   PTI-001  and from  Spanish  project  PGC2018-094792-B-100(MCIU/AEI/FEDER, EU). All authors acknowledge financial support from Proyecto Sinérgico CAM 2020 Y2020/TCS-6545
(NanoQuCo-CM).

	\let\oldaddcontentsline\addcontentsline
	\renewcommand{\addcontentsline}[3]{}
	\bibliographystyle{mybibstylev2}
	\bibliography{library-HAE,Books,booksCarlos,RefsVirtual}

\let\addcontentsline\oldaddcontentsline

\clearpage
\onecolumngrid

\begin{center}
{\bf \large Supplementary Material}
\end{center}

\renewcommand{\theequation}{S\arabic{equation}}

\renewcommand{\thefigure}{S\arabic{figure}} 
\setcounter{figure}{0} 
\setcounter{equation}{0}   

\section{Further details on the {Hierarchical Adiabatic Elimination} }

This section elaborates all the steps involved in the hierarchical adiabatic elimination technique presented in the main text. Let us remind that, as an example, we consider a Hilbert space of three states, $\mathcal{H}=\left\{\ket{1},\ket{2},\ket{V}\right\}$. Here, $|1\rangle$ and $|2\rangle$ represent the ``real'' states, and $|V\rangle$ is an off-resonant state that plays the role of a virtual state that mediates the interactions between $|1\rangle$ and $|2\rangle$. The evolution of the elements of the density matrix are given by the master equation of Eq.~(3) in the main text, 

\begin{subequations}
		\begin{align}
\label{eq:drho_VV}
			\dot \rho_{V,V}&=
			2 \Omega \Im\left[ \rho_{1,V}-\rho_{V,2}\right], \\
			\dot \rho_{2,2}&=-\Gamma \rho_{2,2}+2\Omega \Im \left[\rho_{V,2}\right], \\
			\dot \rho_{1,V}&=i \Delta_V \rho_{1,V}
			 +i\Omega \left[1+\rho_{1,2}-2\rho_{V,V}-\rho_{2,2}\right], \\
			\dot \rho_{1,2}&=-(\Gamma/2)\rho_{1,2}+i\Omega \left[\rho_{1,V}-\rho_{V,2}\right], \\
			\dot \rho_{V,2}&=-\left(i \Delta_V+\Gamma/2\right)\rho_{V,2}			-i\Omega \left[\rho_{1,2}+\rho_{2,2}-\rho_{V,V}\right].
		\end{align}
\end{subequations}
\subsection{First adiabatic elimination}
The first stage of the dynamics can be completely described within the real Hilbert subspace, $\mathcal{H}_{R}=\left\{\ket{1},\ket{2}\right\}$ after an adiabatic elimination of the virtual state. More specifically, this adiabatic elimination consists in setting $\dot{\rho}_{1,V}=\dot{\rho}_{V,2}=0$. This is done under the assumption that the energy difference between the real and virtual subspaces is much larger than the coupling rate, i.e. $\Delta_V \gg \Omega$. 
Thus, by substituting  the virtual coherence terms by their steady state values, the system gets described by the following differential equations
\begin{subequations}
	\begin{align}
		\label{eq:FullEq}
		\dot \rho_{1,2}&\approx\left(-\frac{\Gamma}{2}+\frac{\Gamma \Omega^2}{(i\Gamma -2\Delta_V)\Delta_V}\right)\rho_{1,2}(t)-\frac{i\Omega^2}{\Delta_V}+\left(-\frac{2\Omega^2}{\Gamma+2i\Delta_V}+\frac{i\Omega^2}{\Delta_V}\right)\rho_{2,2}(t)+\left( \frac{2\Omega^2}{\Gamma+2i\Delta_V}+\frac{2i\Omega^2}{\Delta_V} \right)\rho_{V,V}(t), \\
		\dot \rho_{2,2}&\approx\left(-\Gamma -\frac{4 \Gamma \Omega^2 }{\Gamma^2+4\Delta_V^2} \right) \rho_{2,2}(t)-\frac{2\Omega^2}{\Gamma+2 i \Delta_V}\rho_{1,2}(t)-\frac{2\Omega^2}{\Gamma-2 i \Delta_V}\rho_{2,1}(t)+\frac{4\Gamma \Omega^2}{\Gamma^2+4\Delta_V^2}\rho_{V,V}(t).
	\end{align}
\end{subequations}
In the limit $\Delta_V \gg \Gamma$, we can simplify them and obtain a more familiar set of equations,
\begin{subequations}
	\begin{align}
		\dot \rho_{1,2}&\approx-\frac{\Gamma}{2} \rho_{1,2}(t)-i\Omega_{2\mathrm p} (1-2\rho_{2,2}(t))+i\Omega_{2\mathrm p} \rho_{V,V}(t), \\
		\dot \rho_{2,2}&\approx-\Gamma \rho_{2,2}(t)-2\Omega_{2\mathrm p}\Im [\rho_{1,2}(t)]+\frac{\Gamma}{\Delta_V}\Omega_{2\mathrm p}\rho_{V,V}(t).
	\end{align}
			\label{eq:SimplEq}
\end{subequations}
These formulas show a well-known structure, since they correspond to two resonant levels coupled via a second-order process with a Rabi frequency, $\Omega_{2\mathrm p}\equiv \Omega^2/\Delta_V$ along with a standard decay, $\Gamma$. The only addition is an extra term related to the virtual population, $\rho_{V,V}$. However, this term evolves in a much slower timescale than $\rho_{1,2}$ and $\rho_{2,2}$, so in these equations it can be treated as a time-independent parameter with a fixed value. At the beginning of the evolution, we may set $\rho_{V,V}=0$. The time-dependent analytical solutions of these equations are well known and can be found in any quantum optics textbook,
\begin{subequations}
	\begin{align}
	\rho_{2,2}(t)&\approx\frac{4\Omega_{2\mathrm p}^2}{\Gamma^2+8\Omega_{2\mathrm p}^2} 	\left[1-e^{-3\Gamma t/4}\left(\cosh(\kappa t)+\frac{3\Gamma}{4\kappa}\sinh(\kappa t) \right)  \right], \\
	\rho_{1,2}(t)&\approx  \frac{-2 i \Omega_{2\mathrm p} \Gamma}{\Gamma^2+8\Omega_{2\mathrm p}^2}\left[1-e^{-3\Gamma t/4}\left(\cosh(\kappa t)+\left(\frac{\kappa}{\Gamma}+\frac{3\Gamma}{16\kappa}\right)\sinh(\kappa t) \right)  \right],\\
	\rho_{i,V}(t)&=\rho_{V,V}(t)\approx 0\quad (i=1,2),
\end{align}
\end{subequations}
where $\kappa\equiv \frac{1}{2}\sqrt{\frac{\Gamma^2}{4}-16\Omega_{2\mathrm p}^2}$. One can clearly see that, for these equations, the relaxation time towards a stationary state occurs in a timescale $\sim 1/\Gamma$.

\subsection{Second adiabatic elimination}
In much longer timescales than $1/\Gamma$, $\rho_{2,2}$ and $\rho_{1,2}$ can be considered as ``fast'' variables, since they relax to a steady state in a very short time. This allows us to perform a second adiabatic elimination: from Eqs.~\eqref{eq:SimplEq}, it is clear that, if we assume that $\rho_{V,V}$ will be virtually unchanged in a timescale $\sim 1/\Gamma$, we could take it as a time-independent parameter and obtain a stationary solution for $\rho_{2,2}$ and $\rho_{1,2}$ that is dependent on $\rho_{V,V}$. This quasi-steady state will adiabatically follow any slow change of $\rho_{V,V}$. The expression of this $\rho_{V,V}$-dependent steady state  can be obtained by solving a linear system equations of the form $M.\vec{\rho}+\vec{b}=0$ for the vector $\vec{\rho}=\left\{\rho_{2,2}^{SS},\rho_{2,1}^{SS},\rho_{1,2}^{SS}\right\}$, where $M$ and $\vec b$ are given by
\begin{equation}
	M=	\begin{bmatrix}
		  -\Gamma\left(1+\frac{4 \Omega^2}{\Gamma ^2+4 \Delta_V
			^2}\right) & 
			-\frac{2 \Omega ^2}{\Gamma -2i \Delta_V } &
			 -\frac{2 \Omega ^2}{\Gamma +2i \Delta_V }  \\
		-\frac{2\Omega^2}{\Gamma -2 i\Delta_V }-i\Omega_{2\mathrm p}
 &  
	-\Gamma\left(\frac{1}{2} -\frac{i\Omega_{2\mathrm p}}{\Gamma- 2i \Delta_V} \right)
	  & 0 \\
		-\frac{2\Omega^2}{\Gamma +2 i\Delta_V }+i\Omega_{2\mathrm p} & 0 &   
		-\Gamma \left( \frac{1}{2}-\frac{\Gamma \Omega_{2\mathrm p}}{		i \Gamma -2 \Delta_V }\right)
	\end{bmatrix}
\end{equation}
and 
\begin{equation}
	\vec{b}=
	\begin{pmatrix}
		\frac{4 \Gamma  \Omega ^2}{\Gamma ^2+4\Delta_V ^2} \\ 
		 \frac{2\Omega ^2}{\Gamma -2 i\Delta_V }-2i\Omega_{2\mathrm p}\\ 
      \frac{2\Omega ^2}{\Gamma +2 i\Delta_V }+2i\Omega_{2\mathrm p}
	\end{pmatrix}	\rho _{V,V}(t)+
\begin{pmatrix}
	0 \\ 
i\Omega_{2\mathrm p} \\ 
	-i\Omega_{2\mathrm p}
\end{pmatrix}
.
\end{equation}
By solving this linear system we obtain a set of equations for the quasi-stationary values of $\rho_{2,2}$ and $\rho_{1,2}$ that depend on the population of the virtual state at any given time,
\begin{multline}
\rho_{2,2}^{SS}[\rho_{V,V}(t)]=\frac{16\Omega^4(\Delta_V^2
 		+\Omega^2)}{\Gamma^4 \Delta_V^2+32\Omega^2(\Delta_V^2+\Omega^2)+4\Gamma^2(\Delta_V^4+3\Delta_V^2\Omega^2+\Omega^4)}+\\
 \frac{4\left[\Gamma^2\Delta_V^2\Omega^2-4\Omega^4(\Delta_V^2+\Omega^2)\right]}{\Gamma^4 \Delta_V^2+32\Omega^2(\Delta_V^2+\Omega^2)+4\Gamma^2(\Delta_V^4+3\Delta_V^2\Omega^2+\Omega^4)}\rho_{V,V}(t),
\end{multline}

\begin{multline}
		\rho_{1,2}^{SS}[\rho_{V,V}(t)]=\frac{2 \Omega ^2 \left(-i
 			\Gamma  \Delta_V 
 			\left(\Gamma ^2+4 \Delta_V
 			^2\right)-2 \Gamma  \Omega
 			^2 (\Gamma +4 i \Delta_V )-8
 			\Omega ^4\right)}{4 \Gamma
 			^2 \left(3 \Delta_V ^2 \Omega
 			^2+\Delta_V ^4+\Omega
 			^4\right)+\Gamma ^4 \Delta_V
 			^2+32 \Omega ^4
 			\left(\Delta_V ^2+\Omega
 			^2\right)} +\\
 			\frac{4 \Omega ^2  \left(2 \Gamma  \Omega ^2 (\Gamma +4 i
 			\Delta_V )+\Gamma  \Delta_V  (\Gamma +i \Delta_V ) (2 \Delta +i \Gamma )+12
 			\Omega ^4\right)}{4 \Gamma ^2 \left(3 \Delta_V ^2 \Omega ^2+\Delta_V
 			^4+\Omega ^4\right)+\Gamma ^4 \Delta_V ^2+32 \Omega ^4 \left(\Delta_V
 			^2+\Omega ^2\right)}\rho_{V,V}(t).
\end{multline}

After the first adiabatic elimination, the differential equation that governs the dynamics of $\rho_{V,V}$ became:
\begin{equation}
\dot \rho_{V,V}(t) \approx -\frac{4\Gamma \Omega^2}{\Gamma^2 + 4\Delta_V^2} \rho_{V,V} +
 \frac{\Omega_{2\mathrm p}\Gamma(i\Gamma+2\Delta_V)}{\Gamma^2 + 4\Delta_V^2} \rho_{1,2} +
 \frac{\Omega_{2\mathrm p}\Gamma(-i\Gamma+2\Delta_V)}{\Gamma^2 + 4\Delta_V^2} \rho_{2,1} + \frac{4\Gamma}{\Gamma^2 + 4\Delta^2}\rho_{2,2}.
\end{equation}
Substituting the pseudo-stationary values of $\rho_{1,2}$ and $\rho_{2,2}$ into this equation, we obtain a differential equation for $\rho_{V,V}(t)$ which is function of itself, i.e., $\dot{\rho}_{V,V}(t)=f[\rho_{V,V}(t);\rho_{1,2}^{SS}[\rho_{V,V}(t)];\rho_{2,2}^{SS}[\rho_{V,V}(t)]] $. Namely, the differential equation for the virtual state population becomes:
 \small
\begin{equation}
	\dot{\rho}_{V,V}(t)=\frac{4\Gamma \Omega^4(\Gamma^2+4\Omega^2)}{\Gamma^4 \Delta_V^2+32\Omega^2(\Delta_V^2+\Omega^2)+4\Gamma^2(\Delta_V^4+3\Delta_V^2\Omega^2+\Omega^4)} - \frac{4\left[12\Gamma\Omega^6+
		\Gamma^3\Omega^2(\Delta_V^2+2\Omega^2)\right]}{\Gamma^4 \Delta_V^2+32\Omega^2(\Delta_V^2+\Omega^2)+4\Gamma^2(\Delta_V^4+3\Delta_V^2\Omega^2+\Omega^4)}\rho_{V,V}(t),
\end{equation}
\normalsize
obtaining
\begin{equation}
	\rho_{V,V}(t)=\rho_{V,V}^{SS}\left(1-e^{-\Gamma_c t}\right),
\end{equation}
where $\rho_{V,V}^{SS}$ stands for the virtual steady state population (its expression will be given in the next section) and $\Gamma_c$ stands for the relaxation rate, which corresponds to the Liouvillian gap:
\begin{equation}
	\Gamma_c= \frac{4\left[12\Gamma\Omega^6+
		\Gamma^3\Omega^2(\Delta_V^2+2\Omega^2)\right]}{\Gamma^4 \Delta_V^2+32\Omega^2(\Delta_V^2+\Omega^2)+4\Gamma^2(\Delta_V^4+3\Delta_V^2\Omega^2+\Omega^4)}.
\end{equation}
We can reduce this formula under the assumption $\Delta_V \gg \Omega, \Gamma$ and the effective strong coupling regime $\Omega_{2\mathrm p} \gg \Gamma$. In this situation, the relaxation rate reduces to
\begin{equation}
	\Gamma_c\approx \frac{3 \Gamma \Omega^2}{2\Delta_V^2}.
\end{equation}
\subsection{Summary of analytic expressions for the time-dependent density matrix elements}

Once we know the analytic expression for the time-dependent virtual state population, the remaining formulas are easily computed:
\small
\begin{subequations}
	\begin{align}
		\rho_{V,V}(t)&\approx \rho_{V,V}^{SS} \left[1-e^{-\Gamma_c t}\right] ,  \\
		\rho_{2,2}(t)&\approx \rho_{2,2}^{SS} \left[1+\frac{4 \Omega ^4 \left(\Gamma ^2+4 \Delta_V ^2\right)-\Gamma ^4 \Delta_V
			^2+16 \Omega ^6}{4 \Gamma ^2 \left(3 \Delta_V ^2 \Omega ^2+\Delta_V
			^4+\Omega ^4\right)+\Gamma ^4 \Delta_V ^2+32 \Omega ^4 \left(\Delta_V
			^2+\Omega ^2\right)}e^{-\Gamma_c t}\right],    \\
		\rho_{1,2}(t)&\approx \rho_{1,2}^{SS}  \left[1-\frac{2 i \Omega ^2 \left(\Gamma ^2+4 \Omega ^2\right) \left(2 \Gamma 
			\Omega ^2 (\Gamma +4 i \Delta_V )+\Gamma  \Delta_V  (\Gamma +i \Delta_V )
			(2 \Delta_V +i \Gamma )+12 \Omega ^4\right)}{\Gamma  \Delta_V  \left(4
			\Gamma ^2 \left(3 \Delta_V ^2 \Omega ^2+\Delta_V ^4+\Omega
			^4\right)+\Gamma ^4 \Delta_V ^2+32 \Omega ^4 \left(\Delta_V ^2+\Omega
			^2\right)\right)}e^{-\Gamma_c t}\right] ,  \\
		\rho_{1,V}(t)&\approx \rho_{1,V}^{SS}  \left[1+\frac{2 \Omega ^2 \left(\Gamma ^2+4 \Omega ^2\right) \left(-2 i \Gamma 
			\Omega ^2 \left(\Gamma ^2+6 i \Gamma  \Delta_V +2 \Delta_V
			^2\right)+\Gamma ^2 \Delta_V  \left(\Gamma ^2+4 \Delta_V ^2\right)+8
			\Omega ^4 (3 \Delta_V -2 i \Gamma )\right)}{\Gamma  \left(\Gamma 
			\Delta_V -2 i \Omega ^2\right) \left(4 \Gamma ^2 \left(3 \Delta_V ^2
			\Omega ^2+\Delta_V ^4+\Omega ^4\right)+\Gamma ^4 \Delta_V ^2+32 \Omega ^4
			\left(\Delta_V ^2+\Omega ^2\right)\right)}e^{-\Gamma_c t}\right] ,  \\
		\rho_{2,V}(t)&\approx \rho_{2,V}^{SS}  \left[1-\frac{\left(\Gamma ^2+4 \Omega ^2\right) \left(\Gamma ^2 \Delta_V ^2
			(\Gamma +2 i \Delta_V )-4 \Omega ^4 (\Gamma -6 i \Delta_V )+4 i \Gamma 
			\Delta_V  \Omega ^2 (\Gamma +i \Delta_V )\right)}{4 \Gamma ^3 \left(3
			\Delta_V ^2 \Omega ^2+\Delta_V ^4+\Omega ^4\right)+\Gamma ^5 \Delta_V ^2+32
			\Gamma  \Omega ^4 \left(\Delta_V ^2+\Omega ^2\right)}e^{-\Gamma_c t}\right],   \\
	\end{align}
\end{subequations}
\normalsize
where $\rho_{i,j}^{SS}$ $(i,j=1,2,V)$ are the steady state density matrix elements,
\begin{subequations}
	\begin{align}
	\rho_{V,V}^{SS}&=\frac{\Omega ^2 \left(\Gamma ^2+4 \Omega ^2\right)}{\Gamma ^2
		\left(\Delta_V ^2+2 \Omega ^2\right)+12 \Omega ^4} ,\qquad 
	\rho_{2,2}^{SS}=\frac{4 \Omega ^4}{\Gamma ^2 \left(\Delta_V ^2+2 \Omega ^2\right)+12
		\Omega ^4} ,\\
	\rho_{1,2}^{SS}&=-\frac{2 i \Gamma  \Delta_V  \Omega ^2}{\Gamma ^2 \left(\Delta_V ^2+2 \Omega
		^2\right)+12 \Omega ^4} , \qquad
	\rho_{1,V}^{SS}=\frac{\Gamma  \Omega  \left(-\Gamma  \Delta_V +2 i \Omega
		^2\right)}{\Gamma ^2 \left(\Delta_V ^2+2 \Omega ^2\right)+12 \Omega ^4} ,\qquad 
	\rho_{2,V}^{SS}= -\frac{2 i \Gamma  \Omega ^3}{\Gamma ^2 \left(\Delta_V ^2+2 \Omega
		^2\right)+12 \Omega ^4}.\\
	\end{align}
\end{subequations}
The analytic expression within the metastability regime are just the one for a single two-level system,
\begin{equation}
	\label{eq:TwoPhotonSS}
		\rho_{2,2}^{M}=\frac{4\Omega_{2\mathrm p}^2}{\Gamma^2+8\Omega_{2\mathrm p}^2} , \quad
		\rho_{1,2}^{M}=\frac{2i\Gamma \Omega_{2\mathrm p}}{\Gamma^2+8\Omega_{2\mathrm p}^2} , \quad \rho_{i,V}^{M}\approx 0 \quad \forall i=1,2,V.
\end{equation}

\section{Calculation of the concurrence}
We now consider that the Hilbert space $\mathcal{H}=\left\{\ket{1},\ket{2},\ket{V}\right\}$ is mapping the Hilbert space of two interacting quantum emitters, in such a way that $\left\{|1\rangle, |2\rangle, |V\rangle \right\}$ now stand for the triplet $\left\{|gg \rangle, |ee\rangle,\ket{S}= 1/\sqrt{2}(|ge \rangle +|eg \rangle\right\}$, where the antisymmetric state is neglected due to be completely disconnected from the dynamics. In this scenario, we make the assumption that $\rho$ has the following structure
 \begin{equation}
 	\rho \approx \begin{pmatrix}
 			\rho_{1,1} & 0 & 0 & \rho_{1,2} \\
 			0 & \rho_{V,V} & 0 &0 \\
 			0 & 0 &0&0\\
 			\rho_{2,1} &0&0&\rho_{2,2}
 	\end{pmatrix},
 \end{equation}
so that the square roots of the eigenvalues of the density matrix  $\rho \sigma_y \otimes \sigma_y \rho^* \sigma_y \otimes \sigma_y$ are given by
	\begin{align}
		\lambda_1&=\sqrt{\left| \rho _{1,2}\right| {}^2+2 \left| \rho _{1,2}\right|  \sqrt{-\rho _{2,2} \left(\rho _{2,2}+\rho _{V,V}-1\right)}-\rho _{2,2} \left(\rho _{2,2}+\rho _{V,V}-1\right)},\\
		\lambda_2&=\sqrt{\left| \rho _{1,2}\right| {}^2-2 \left| \rho _{1,2}\right|  \sqrt{-\rho _{2,2} \left(\rho _{2,2}+\rho _{V,V}-1\right)}-\rho _{2,2} \left(\rho _{2,2}+\rho _{V,V}-1\right)}\\
		\lambda_3&=\rho_{V,V}.
	\end{align}
This expressions allows us to compute analytically the expression for the concurrence. This can be done taking into account the analytic formulas shown before,  Eq.~\eqref{eq:TwoPhotonSS}. The metastable value of non-zero concurrence that is achieved is thus given by:
	\begin{equation}
		\mathcal{C}^{\text{m}}(\rho)=\frac{2\sqrt{2}|\Omega_{2\mathrm p}|}{\Gamma^2 +8|\Omega_{2\mathrm p}|^2}\left(
		\sqrt{ 2|\Omega_{2\mathrm p}|^2+\Gamma^2+\Gamma \sqrt{\Gamma^2+4|\Omega_{2\mathrm p}|^2}  }
		-\sqrt{ 2|\Omega_{2\mathrm p}|^2+\Gamma^2-\Gamma \sqrt{\Gamma^2+4|\Omega_{2\mathrm p}|^2}  }
		\right).
	\end{equation}
This expression allows us to establish the optimum value of $|\Omega_{2\mathrm p}|$ that maximizes the metastable value of the concurrence,
\begin{equation}
	|\Omega_{2\mathrm p}|^{opt}=\frac{\Gamma}{2\sqrt{2}}.
\end{equation}

\end{document}